\documentclass{article} 
\usepackage{a4wide,amssymb} 
\usepackage{graphics}

\newcommand{\vc}{\vec}
\newcommand{\rund}[1]{\left(#1\right)}
\newcommand{\abs}[1]{\left|#1\right|}
\newcommand{\wave}[1]{\left\{#1\right\}}
\newcommand{\eck}[1]{\left[#1\right]}
\newcommand{\ave}[1]{\left\langle#1\right\rangle}
\newcommand{\eps}{\epsilon}
\newcommand{\eckk}{\eck}
\newcommand{\Real}{\Re}
\newcommand{\D}{{\cal D}}
\newcommand{\U}{{\cal U}}
\newcommand{\R}{{\cal R}}
\newcommand{\s}{{\rm s}}
\newcommand{\arcminf}{\farcm}
\renewcommand{\d}{{\rm d}}
\renewcommand{\L}{{\cal L}}
\newlength{\asize}
\newlength{\bsize}
\newlength{\csize}
\newcommand{\farcm}{\hbox{$.\!\!^{\prime}$}} 

\begin{document}



\title{Entropy-regularized Maximum-Likelihood cluster 
  mass reconstruction} 
\author{Stella Seitz$^{1,2}$, Peter Schneider$^{2}$ and 
  Matthias Bartelmann$^{2}$ \and 
  $^1$ Universit\"atssternwarte M\"unchen, Scheinerstr.~1, 
  D-81679 M\"unchen, Germany \and 
  $^2$ Max-Planck-Institut f. Astrophysik, Postfach 1523, \\ 
  D-85740 Garching, Germany} 


\date{} 

\maketitle

\begin{abstract}
We present a new method for reconstructing two-dimensional mass maps
of galaxy clusters from the image distortion of background
galaxies. In contrast to most previous approaches, which directly
convert locally averaged image ellipticities to mass maps (direct
methods), our entropy-regularized maximum-likelihood method is an
inverse approach. Albeit somewhat more expensive computationally, our
method allows high spatial resolution in those parts of the cluster
where the lensing signal is strong enough. Furthermore, it allows to
straightforwardly incorporate additional constraints, such as
magnification information or strong-lensing features. Using synthetic
data, we compare our new approach to direct methods and find indeed a
substantial improvement especially in the reconstruction of mass
peaks. The main differences to previously published inverse methods
are discussed.


\end{abstract}

\section{Introduction}

The reconstruction of projected cluster mass maps from the observable
image distortion of faint background galaxies due to the tidal
gravitational field is a new and powerful technique. Pioneered by
Kaiser \& Squires (1993), this method has since been modified and
generalized to account for (a) strong tidal fields in cluster centers
(Schneider \& Seitz 1995; Seitz \& Schneider 1995; Kaiser 1995); (b)
finite and -- in some cases, e.g.~WFPC2 images -- very small data
fields (Schneider 1995; Kaiser et al.\ 1995; Bartelmann 1995; Seitz
\& Schneider 1996, 1998; Lombardi \& Bertin 1998); and (c) the broad
redshift distribution of background galaxies (Seitz \& Schneider
1997). All of these are direct methods in the sense that a local
estimate of the tidal field is derived from observed galaxy
ellipticities, which is then inserted into an inversion equation to
obtain an estimate of the surface mass density of the cluster.

Whereas these direct methods are computationally fast, can be treated
as black-box routines, need only the observed ellipticities and a
smoothing length $\theta_0$ as input data, and yield fair estimates of
the surface mass density, their application has several drawbacks:

\begin{itemize}

\item The data must be smoothed, and the smoothing scale is typically
  a free input parameter specified prior to the mass reconstruction.
  There are no objective criteria on how to set the smoothing scale,
  although some ad-hoc prescriptions for adapting it to the strength
  of the lensing signal have been given (Seitz et al.\ 1996). In
  general, smoothing leads to an underestimate of the surface mass
  density in cluster centers or sub-condensations.

\item The quality of the reconstruction is hard to quantify.

\item Constraints on the mass distribution from additional observables
  (such as multiple images or giant arcs) cannot simultaneously be
  included. In particular, magnification information contained in the
  number density of background sources (Broadhurst et al.\ 1995; Fort
  et al.\ 1997) or in the image sizes at fixed surface brightness
  (Bartelmann \& Narayan 1995), cannot be incorporated locally but
  only globally to break the mass-sheet degeneracy (Gorenstein et al.\
  1988; Schneider \& Seitz 1995).

\end{itemize}

To overcome these drawbacks, a different class of methods should be
used. Bartelmann et al.\ (1996, hereafter BNSS) developed a
maximum-likelihood (ML) technique in which the values of the
deflection potential at grid points are considered as free
parameters. After averaging image ellipticities and sizes over grid
cells, local estimates of shear and magnification are obtained. The
deflection potential at the grid points is then determined such as to
optimally reproduce the observed shear and magnification estimates.
Magnification information can be included this way. The smoothing
scale in this method is given by the size of the grid cells, and can
be chosen such that the overall $\chi^2$ of the fit is of order unity
per degree of freedom.

Squires \& Kaiser (1996; hereafter SK) suggested several inverse
methods. Their {\em maximum probability method\/} parameterizes the
mass distribution of the cluster by a set of Fourier modes. If the
number of degrees of freedom (here the number of Fourier modes) is
large, the mass model tends to over-fit the data. This has to be
avoided by regularizing the model, for which purpose SK impose a
condition on the power spectrum of the Fourier modes. SK's {\em
maximum-likelihood method\/} specifies the surface mass density on a
grid and uses the Tikhonov-Miller regularization (Press et al.\ 1992,
Sect.\ 18.5). The smoothness of the mass reconstructions can be
changed by varying the regularization parameter, which is chosen such
as to give an overall $\chi^2\approx1$ per degree of freedom.

Bridle et al.\ (1998) have recently proposed an entropy-regularized ML
method in which the cluster mass map is parameterized by the surface
mass density at grid points. This method allows to restrict the
possible mass maps to such with non-negative surface mass density.

This paper describes another variant of the ML method (Seitz 1997,
Ph.D.\ thesis). The major differences to the previously mentioned
inverse methods are the following:

\begin{itemize}

\item The observational data (e.g.~the image ellipticities) are not
  smoothed, but each individual ellipticity of a background galaxy is
  used in the likelihood function. Whereas this modification
  complicates the implementation of the method, it allows larger
  spatial resolution for a given number of grid points, which is
  useful since the latter determines the computing time.

\item The number of grid points can be much larger than in BNSS, and
  the likelihood function is regularized. This produces mass
  reconstructions of variable smoothness: Mass maps are smooth where
  the data do not demand structure, but show sharp peaks where
  required by the data. The resulting spatially varying smoothing
  scale is a very desirable feature. Fourier methods, such as SK's
  maximum probability method, have a spatially constant smoothing
  scale which is determined by the highest-order Fourier components.
  They always need to compromise between providing sufficient
  resolution near mass peaks and avoiding over-fitting of the data in
  the outer parts of a cluster.

\item Following BNSS, we use the deflection potential to describe a
  cluster. This is an essential difference to Bridle et al.\ (1998)
  who used the surface mass density at grid points. As we shall
  discuss below, working with the deflection potential has substantial
  fundamental and practical advantages.

\end{itemize}

We describe our method in Sect.\ 2, with details given in the
Appendix. We then apply the method to synthetic data sets in Sects.\ 3
\& 4 to demonstrate its accuracy. In particular, we compare the
performance of our ML method to that of direct methods. The results
are then discussed in Sect.\ 4, and conclusions are given in Sect.\ 5,
where we also discuss further generalizations of the method for, e.g.,
including constraints from strong lensing features.

\section{The entropy-regularized ML mass reconstruction}

\subsection{Basic lensing relations}

For simplicity, we assume throughout the paper that all background
galaxies are located at the same redshift. A generalization of our
technique to a redshift distribution is given by Geiger \& Schneider
(1998). The dimensionless surface mass density $\kappa(\vc\theta)$ is
related to the deflection potential $\psi(\vc\theta)$ through the
Poisson equation,
\begin{equation}
  \kappa(\vc\theta) = {1\over 2}\rund{\psi_{,11}+\psi_{,22}}\;,
\label{eq:2.1}
\end{equation}
where indices $i$ preceded by a comma denote partial derivatives with
respect to $\theta_i$. The tidal gravitational field of the lens is
described by the shear $\gamma=\gamma_1+{\rm i}\,\gamma_2$ with the
two components
\begin{equation}
  \gamma_1 = {1\over 2}\rund{\psi_{,11}-\psi_{,22}}\;,\quad
  \gamma_2 = \psi_{,12}\;.
\label{eq:2.2}
\end{equation}
Thus, the surface mass density and the shear, which determine the
local properties of the lens mapping, can {\em locally\/} be obtained
from the deflection potential $\psi$. In contrast, the relation
between shear and surface mass density is highly non-local,
\begin{equation}
  \gamma(\vc\theta) = {1\over\pi}\,
  \int_{\Real^2}\d^2\theta'\;
  \D(\vc\theta-\vc\theta')\,\kappa(\vc\theta')\;,
\label{eq:2.3}
\end{equation}
with $\D(\vc\theta) = -(\theta_1^2 - \theta_2^2 + 2\,{\rm
i}\,\theta_1\theta_2)|\vc\theta|^{-4}$. In particular, $\kappa$ needs
to be given on the entire two-dimensional plane. Prescribing $\kappa$
on a finite field does therefore not completely specify the shear
inside the field, because the latter is also affected by the outside
mass distribution. We return to this point further below. The local
magnification is
\begin{equation}
  \mu(\vc\theta) = \wave{
    \eck{1-\kappa(\vc\theta)}^2-\abs{\gamma(\vc\theta)}^2
  }^{-1}\;.
\label{eq:2.4}
\end{equation}

The local lens equation relates the ellipticities of a source and its
image. We use the complex ellipticity parameter $\chi$ (see Blandford
et al.\ 1991) to describe image shapes. It is generally defined in
terms of the tensor $Q_{ij}$ of second brightness moments of an image
by $\chi=(Q_{11}-Q_{22}+2\,{\rm i}\,Q_{12})/(Q_{11}+Q_{22})$.

\subsection{The ML method}

We refer the reader to Press et al.\ (1992, Chap.\ 18) for the basic
ingredients of the ML method; see also Bridle et al.\ (1998). We do
not repeat the basics here, but describe the application of the method
to the cluster mass reconstruction. We start by considering image
ellipticities only; magnification effects will be discussed later.

Let $\chi_k$, $1\le k\le N_{\rm g}$, denote the complex ellipticities
of $N_{\rm g}$ galaxy images in the data field $\U$, which we assume
to be a rectangle of side lengths $L_x$ and $L_y$. We cover the data
field with an equidistant grid of $N_x\times N_y$ points
$\vc\theta_{ij}$, with $\vc\theta_{11}$ in the lower left corner of
the data field. The cluster is described by the deflection potential
at the grid points, $\psi_{ij}$, $0\le i\le N_x+1$, $0\le j\le
N_y+1$. As discussed in BNSS, the grid for $\psi$ is larger than the
data field by one column or row of grid points in all four directions
to allow simple finite differencing of $\psi$ on the whole field
$\U$. Having found $\kappa$ and $\gamma$ on all grid points from
$\psi$ according to (\ref{eq:2.1}) and (\ref{eq:2.2}), $\kappa$ and
$\gamma$ are bilinearly interpolated to all galaxy positions
$\vc\theta_k$.

If the isotropic probability distribution $p_{\rm s}(\chi^\s)$ of the
intrinsic source ellipticities is given, the probability distribution
$p(\chi;\wave{\psi})$ of the image ellipticities can be predicted.
The likelihood function is then defined as
\begin{equation}
  \L(\wave{\psi}) := \prod_{k=1}^{N_{\rm g}}\,
  p(\chi_k;\wave{\psi})\;.
\label{eq:2.5}
\end{equation}
$\L$, or $\ln\L$, can be maximized with respect to the set
$\wave{\psi}$ of values of the deflection potential at the grid
points. Since the values of $\kappa$ and $\gamma$ are obtained from
second derivatives of $\psi$, a constant and a term linear in
$\vc\theta$ added to $\psi$ leave $\L$ unchanged. In addition, the
mass-sheet degeneracy renders $\L$ invariant under the transformation
\begin{equation}
  \psi(\vc\theta) \to \lambda\,\psi(\vc\theta) +
  {(1-\lambda)\over2}\,|\vc\theta|^2 \;,
\label{eq:2.6}
\end{equation}
where $\lambda\ne0$ is an arbitrary parameter (Schneider \& Seitz
1995). Therefore, in maximizing $\L$ with respect to $\wave{\psi}$,
the potential $\psi$ can be held fixed at four grid points. Noting that
the corners of the $\psi$ grid are not used for the calculation of
$\kappa$ and $\gamma$ on $\U$ (see Appendix), we see that the
maximization of $\L$ has dimension $N_{\rm dim}=(N_x+2)(N_y+2)-8$.

\subsection{Regularization}

Provided $N_{\rm dim}$ is not much smaller than the number of galaxies
(which we assume in the following), the maximization results in a
cluster model which tries to follow closely the noise pattern of the
data. Disregarding observational effects, the noise is due to the
intrinsic ellipticity distribution of the sources. The reconstructed
mass distribution will therefore have pronounced small-scale
structure, fitting the observed image ellipticities as closely as
possible, and having a $\chi^2$ per degree of freedom much smaller
than unity. In order to prevent such over-fitting of the data, we need
to augment $\L$ by a regularization term. Instead of maximizing
$\ln\L$, we minimize
\begin{equation}
  E\rund{\wave{\psi}} := -{1\over N_{\rm g}}\ln\L({\wave{\psi}}) +
  \eta\,\R({\wave{\psi}})\;, 
\label{eq:2.7}
\end{equation}
where $\R$ is a function of the potential that disfavors strong
small-scale fluctuations. The parameter $\eta$ determines how much
weight should be attached to smoothness. One can vary $\eta$ such that
the resulting reconstruction has approximately the expected deviation
from the data, viz.~$\chi^2\approx1$ per degree of freedom. Larger
values of $\eta$ yield mass distributions which are too smooth to fit
the data, lower values of $\eta$ cause over-fitting.

We experimented with quite a large number of regularization terms. For
example, we chose $\R$ as the sum of $|\nabla\kappa|^2$ over all grid
points. Mass reconstructions from synthetic data (see Sect.\ 3 below)
then showed a strong tendency to decrease too slowly towards the outer
parts of the cluster, for that regularization preferred $\kappa$ to be
as flat as possible. Regularizations including higher-order
derivatives of $\kappa$ (see Press et al.\ 1992, Sect.\ 18.5) led to
similar artifacts. Thus, such local linear regularizations were
dismissed as unsatisfactory.

Motivated by the success of the maximum-entropy (ME) image
deconvolution (e.g.~Narayan \& Nityananda 1986; Lucy 1994), we
consider instead ME regularizations of the form
\begin{equation}
  \R = \sum_{i,j=1}^{{N_x},{N_y}}\,\hat\kappa_{ij}\,
  \ln\rund{\hat\kappa_{ij}\over b_{ij}}\;,
\label{eq:2.8}
\end{equation}
where 
\begin{equation}
  \hat\kappa_{ij} = \eck{\sum_{k,l=1}^{{N_x},{N_y}}\,
  \kappa_{kl}}^{-1}\,\kappa_{ij}
\label{eq:2.9}
\end{equation}
is the normalized surface mass density at the grid points, and
$b_{ij}$ is a similarly normalized prior distribution (see Press et
al.\ 1992, Sect.\ 18.7, for a detailed discussion of the ME
method). We experimented with different choices for the prior. On the
whole, a uniform prior, $b_{ij}={\rm const.}$, performed
satisfactorily, but tended to smooth mass peaks more than
desired. Following Lucy (1994), we therefore use a prior which is
determined by the data itself. Deferring details to Sect.\ 3, we note
here that one can use the mass distribution obtained from a direct
(finite field) reconstruction method as an initial prior, then
iteratively minimize $E$, and after several iterations use a smoothed
version of the current mass distribution as a new prior. Lucy (1994)
showed that such a moving prior yields more accurate reconstructions
than a constant prior. The regularization parameter $\eta$ can be
iteratively adjusted to provide the expected goodness-of-fit. Note
that the ME regularization ensures that the reconstructed surface mass
distribution is positive definite.

\subsection{The ellipticity distribution}

Given the intrinsic ellipticity distribution and the local values of
$\kappa$ and $\gamma$, the probability distribution $p(\chi)$ for the
image ellipticities can be calculated. However, the resulting analytic
expressions are quite cumbersome and unsuitable for the
high-dimensional minimization problem considered here. With the
intrinsic distribution not being accurately known anyway, a precise
expression of $p(\chi)$ is not needed. We therefore approximate the
image ellipticity distribution by a Gaussian, with mean $\ave{\chi}$
and dispersion $\sigma_\chi$. Both these values depend on the (local)
distortion,
\begin{equation}
  \delta = {2g\over 1 + \abs{g}^2}\;,
\label{eq:2.10}
\end{equation}
where $g=\gamma/(1-\kappa)$ is the reduced shear. Mean and dispersion
can be approximated by (Schneider \& Seitz 1995)
\begin{eqnarray}
  \ave{\chi} &=& \zeta\,\delta \equiv \eck{1-{M_2\over2}\,
  \rund{1-\abs{\delta}^2}^{\mu_1}}\delta\;,\nonumber\\
  \sigma_\chi &=& \sigma_0\rund{1-\abs{\delta}^2}^{\mu_2}\;,
\label{eq:2.11}
\end{eqnarray}
where
\begin{equation}
  \mu_1 = 1-{3M_4\over 4 M_2}\;,\quad
  \mu_2 = {6M_2+M_2^2-9M_4\over8M_2}\;,
\label{eq:2.12}
\end{equation}
and $M_n$ is the $n$-th moment of the intrinsic ellipticity
distribution. Using (\ref{eq:2.5}) and (\ref{eq:2.7}), the function
$E$ to be minimized can then be written
\begin{equation}
  E\rund{\wave{\psi}} = G_\chi\rund{\wave{\psi}} +
  \sum_{k=1}^{N_{\rm g}}\,\ln\eck{\sigma^2_\chi(k)} +
  \eta\,\R\rund{\wave{\psi}}\;,
\label{eq:2.13}
\end{equation}
where we have introduced
\begin{equation}
  G_\chi := {1\over N_{\rm g}}\sum_{k=1}^{N_{\rm g}}
  {\abs{\chi_k-\ave{\chi}(k)}^2 \over \sigma^2_\chi(k)}\;.
\label{eq:2.14}
\end{equation}
$\ave{\chi}(k)$ and $\sigma_\chi(k)$ are the values of $\ave{\chi}$
and $\sigma_\chi$ at the position of the $k$-th galaxy, which
obviously depend on the deflection potential $\wave{\psi}$ through the
distortion $\delta$.\footnote{If we worked in terms of the ellipticity
parameter $\eps$ (see BBNS) rather than the ellipticity $\chi$, then
$\ave{\eps}=g$ (Seitz \& Schneider 1997) for a non-critical
cluster. In the general case including critical clusters, the variable
$\chi$ is more convenient.} The term $G_\chi$ in $E$ has the form of a
$\chi^2$-function, which implies that an acceptable mass model should
have $G_\chi\approx 1$. This condition constrains the regularization
parameter $\eta$.

We outline in the Appendix an efficient method for calculating $E$ and
its derivatives with respect to the values of $\psi$ on the grid
points. We note that the terms in the sums of (\ref{eq:2.13}) and
(\ref{eq:2.14}) corresponding to a galaxy at $\vc\theta_{k}$ depend
only on the values of $\psi$ at neighboring grid points. In that
sense, our method is local. Had we parameterized the cluster by
$\kappa$ at grid points, each term in the sums depended on $\kappa$ at
all grid points, as can be seen from (\ref{eq:2.3}). It becomes
obvious that the description in terms of $\psi$ requires much less
computer time. In addition, the use of $\kappa$ rather than $\psi$ is
strongly disfavored by the fact that the shear on $\U$ is
incompletely specified by $\kappa$ on the same field. Bridle et al.\
(1998) attack this problem by performing the reconstruction on a
region much larger than $\U$. They find that the shear information
within $\U$ yields information about $\kappa$ outside $\U$. Although
this is true, the information on the mass outside the data field is
{\em very\/} limited: For a circular data field, the shear inside $\U$
caused by mass outside $\U$ can be fully described by conveniently
defined multipole moments of the mass distribution outside $\U$
(Schneider \& Bartelmann 1997), and there are infinitely many mass
distributions for which all of those multipole moments agree. For
instance, a point mass located just outside the data field produces
the same shear pattern in the field as a spherically symmetric mass
distribution with the same total mass. Finally, extending the region
on which the reconstruction is performed increases the dimensionality
of the minimization problem.

\subsection{Including magnification information}

The mass-sheet degeneracy (\ref{eq:2.6}) can be lifted if the lens
magnification can be estimated. Three different methods for measuring
magnification were suggested in the literature. Broadhurst et al.\
(1995) proposed to use the magnification bias, which changes the local
number density of background galaxies due to the magnification,
provided the slope of the number counts $\d\log N/\d m$ is
sufficiently different from $0.4$. Noting that lensing magnifies
objects but leaves their surface brightness unchanged, Bartelmann \&
Narayan (1995) suggested that the sizes of background galaxies at
fixed surface brightness could be a convenient measure for the
magnification after calibrating with field galaxies. Both methods can
be used locally or globally. In the first case, the local
magnification information is used for the mass reconstruction, whereas
in the latter case, the transformation parameter $\lambda$ in
(\ref{eq:2.6}) is adjusted until the magnification optimally matches
the observational estimate. Kolatt \& Bartelmann (1998) suggested to
calibrate $\lambda$ globally by using type-Ia supernovae as
cosmological standard candles. If magnification information is taken
into account, the potential $\psi$ can be kept fixed at three points
only, yielding $N_{\rm dim}=(N_x+2)(N_y+2)-7$.

As an example, consider the method suggested by Bartelmann \& Narayan
(1995). Let $r^\s$ and $r$ be the ratios of the linear sizes of a
galaxy and its image, respectively, relative to the mean size of
galaxies with the same surface brightness. They are related by
$r={\abs{\mu}}^{1/2}\,r^\s$. The expectation values of $r^\s$ and $r$
are $1$ and ${\abs{\mu}}^{1/2}$, respectively. Hence, if $\hat p_{\rm
s}(r^\s)$ is the probability distribution for $r^\s$, the local
probability distribution for $r$ is
\begin{equation}
  \hat p(r) = \abs{\mu}^{-1/2}\,\hat p_{\rm s}
  \rund{\abs{\mu}^{-1/2}\,r}\;.
\label{eq:2.15}
\end{equation}
Including the size distribution into the likelihood maximization leads
to an additional term in (\ref{eq:2.13}),
\begin{eqnarray}
  E\rund{\wave{\psi}} &=& G_\chi\rund{\wave{\psi}} + 
  {1\over 2}G_r\rund{\wave{\psi}} \nonumber\\
  &+& \sum_{k=1}^{N_{\rm g}}\,\ln\eck{\sigma^2_\chi(k)} +
  \eta\,\R\rund{\wave{\psi}}\;,
\label{eq:2.16}
\end{eqnarray}
where
\begin{equation}
  G_r := {2\over N_{\rm g}}\,\sum_{k=1}^{N_{\rm r}}\,
  \abs{\mu(k)}^{-1/2}\hat p_{\rm s}
  \rund{\abs{\mu(k)}^{-1/2}\,r(k)}\; .
\label{eq:2.17}
\end{equation}
We assume $\hat p_{\rm s}$ to be a log-normal distribution
(cf.~Bartelmann \& Narayan 1995),
\begin{equation}
  \hat p_{\rm s}(\ln r^\s) = {1\over\sqrt{2\pi}\sigma_r}\,
  \exp\eck{-{\rund{\ln r^\s+\sigma_r^2/2}^2\over 2\sigma_r^2}}\;,
\label{eq:2.18}
\end{equation}
with $\ave{r^\s}=1$, and (\ref{eq:2.17}) becomes
\begin{equation}
  G_r := {1\over N_{\rm g}}\,\sum_{k=1}^{N_{\rm r}}\,
  {\eckk{\ln r(k)-\ave{\ln r}(k)}^2 \over  \sigma_r^2}\;,
\label{eq:2.19}
\end{equation}
with $\ave{\ln r}(k)=(\ln\abs{\mu(k)}-\sigma_r^2)/2$. Hence, $\hat p$
is a Gaussian in $\ln r$, with dispersion $\sigma_r$. $G_r$ has the
form of a $\chi^2$ function in $\ln r$, which motivates us to include
the factor $1/2$ in the definition of $G_r$. A satisfactory mass model
should have $G_r\approx1$. Note that the galaxies whose size can
reliably be measured need not be those used for measuring the shear.
It is just for notational simplicity that we assume that the two
galaxy populations agree.

\section{Practical implementation and simulation parameters}

\subsection{Practical implementation}

For minimizing (\ref{eq:2.13}) or (\ref{eq:2.16}) in the process of
the ML mass reconstruction, the various quantities
(e.g.~$\ave{\chi}(k)$, $\sigma_\chi(k)$, $\hat\kappa_{ij}$) need to be
calculated for each set of values $\wave{\psi}$ of the deflection
potential. We outline in the Appendix how this can be done
efficiently. In order to quickly approach the minimum of $E$, we use
derivative information in the minimization procedure. The derivative
of $E$ with respect to $\psi_{ij}$ is also given in the Appendix. We
employ the conjugate gradient method as encoded in the routine {\tt
frprmn} by Press et al.\ (1992), with line minimization using
derivative information.

We need a good initial potential to start the minimization. We tested
two different approaches. The first starts with a relatively small
grid (say, $N_x=N_y=11$ for a quadratic data field $\U$), and a
potential which corresponds to a constant surface mass density of
$\kappa\approx 0.1$, say. $\eta$ can be set to zero initially because
the large grid cells provide sufficient smoothing to avoid over-fitting
the data, or otherwise set $\eta$ to a small value with a constant
prior. After several (20, say) iterations, the current mass map is
smoothed and used as the new prior, and the minimization is
continued. The regularization parameter is slightly increased or
decreased, depending on whether $G_\chi$ is smaller or larger than
unity. Once a stable minimum with $G_\chi\approx1$ is obtained, the
solution can be interpolated to a finer grid ($N_x=N_y=21$, say) and
the minimization can continue, adapting the prior and the
regularization parameter as described before. This procedure can be
repeated if desired.

The second approach starts on a fine grid right away, with initial
conditions obtained from a direct finite-field reconstruction method
like that described by Seitz \& Schneider (1998). From this mass
distribution, an approximate deflection potential can be found by
integration, although any contribution from the mass outside the data
field will be missing in the resulting $\psi$. This initial prior is a
smoothed version of the mass map from the direct reconstruction, and
is adapted as described above. If there is no mass concentration
directly outside the data field, the second method converges faster,
while the first approach should be used if the data field does not
encompass most of the mass concentration. Of course, both methods
finally approach solutions with $G_\chi\approx1$. If magnification
information is available, $G_r$ should finally approximate unity,
which provides a useful consistency check of the result.

\begin{figure}[ht]
  \centerline{%
    \resizebox{\asize}{!}{\includegraphics{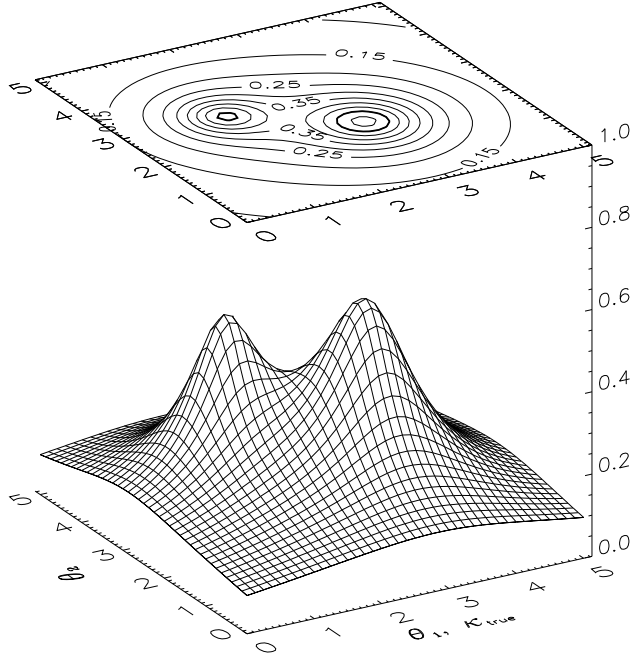}}}
\caption{A mass model consisting of two softened isothermal spheres
  with core sizes $0\arcminf6$ and $0\arcminf8$, and central surface
  mass densities $\kappa_{\rm c}=0.4$ and $0.5$,
  respectively. Contours are spaced by $0.05$, and the heavy contour
  follows $\kappa=0.5$}
\label{fig:1}
\end{figure}

\subsection{Simulation parameters}

We carry out simulations in which background galaxies are lensed by
the mass model shown in Fig.~\ref{fig:1}. It consists of two softened
isothermal spheres, with parameters chosen such that the lens is
sub-critical. We successfully performed simulations with critical
clusters as well, but concentrate on the non-critical case to simplify
comparisons with the noise-filter reconstructions of Seitz \&
Schneider (1996). The modifications to the practical implementation
necessary for critical clusters are given in the Appendix. The data
field $\U$ is a square with side length $5'$. We choose a number
density of 50 galaxies per square arc minute, approximately
corresponding to the number density which can be achieved with several
hours exposure time at a four-meter class telescope in good
seeing. All reconstructions were performed on a grid with
$N_x=N_y=41$. Note that the parameters are chosen such that the grid
cells have about the same size as the mean separation between
galaxies. The intrinsic ellipticity distribution was chosen to be
approximately Gaussian,
\begin{equation}
  p_{\rm s}(\chi^\s) = {1\over \pi R^2[1- {\rm exp}(-1/R^2)]}\,
  {\rm exp}(-\abs{\chi^\s}^2/ R^2)\;,
\label{eq:3.1}
\end{equation}
with $R=0.3$, yielding $M_2\approx 0.09$, $\mu_1\approx 0.87$, and
$\mu_2\approx 0.55$ in (\ref{eq:2.11}) and (\ref{eq:2.12}). The
distribution $\hat p_{\rm s}$ of relative source sizes is
characterized by $\sigma_r=0.5$ (see Bartelmann \& Narayan 1995 for a
discussion of this choice). The ``observed'' ellipticities and
relative source sizes are calculated from realizations of the source
distributions, using the gravitational lens equations.

\begin{figure*}
  \resizebox{\bsize}{!}{\includegraphics{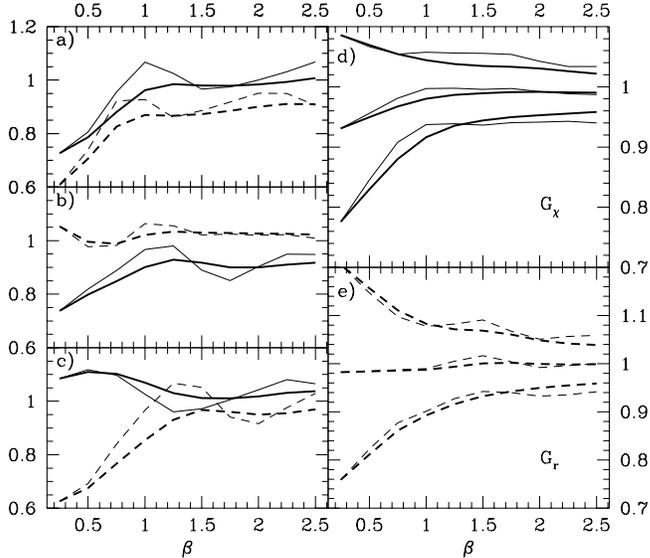}}
  \hfill
  \parbox[b]{\csize}{
\caption{Panels (a) through (c) show the chi-square contributions of
  the observables, viz.\ ellipticity and image size, within circles of
  radius $\beta$, and within rings of outer radius $\beta$ and width
  $\Delta\beta=0\arcminf 25$, for three different realizations of the
  source population. For this figure, the true mass distribution of
  the lens was taken, so that these curves merely show the noise
  properties of the source population. Heavy solid curves:
  $G_\chi(<\beta)$; light solid curves: $G_\chi(\beta)$; heavy dashed
  curves: $G_r(<\beta)$; light dashed curves: $G_r(\beta)$. Panels (d)
  and (e) show the average of these four quantities, obtained from 50
  realizations of the source population, together with their
  1-$\sigma$-variations, shown by the same line types}
\label{fig:2}}
\end{figure*}

In order to assess the expected deviation of $G_\chi$ and $G_r$ from
unity, we plot in Fig.~\ref{fig:2} the quantities
\begin{eqnarray}
  G_\chi(<\beta) &:=& \ave{{\abs{\chi_k-\ave{\chi}(k)}^2 \over 
  \sigma^2_\chi(k)}}_{\abs{\vc \theta_k}<\beta} \;,\nonumber\\
  G_\chi(\beta) &:=& \ave{{\abs{\chi_k-\ave{\chi}(k)}^2 \over 
  \sigma^2_\chi(k)}}_{\beta-\Delta\beta\le\abs{\vc \theta_k}<\beta}
  \;,\nonumber\\
  G_r(<\beta) &:=& \ave{\eckk{\ln r(k)-\ave{\ln
  r}(k)}^2 \over  \sigma_r^2}_{\abs{\vc \theta_k}<\beta}
  \;,\nonumber\\
  G_r(\beta) &:=& \ave{\eckk{\ln r(k)-\ave{\ln
  r}(k)}^2 \over  \sigma_r^2}_{\beta-\Delta\beta
  \le\abs{\vc \theta_k}<\beta} \;.
\label{eq:3.2}
\end{eqnarray}
They are the contributions to $G_\chi$ and $G_r$ from galaxies closer
than $\beta$ to the center of the data field, or within rings of width
$\Delta\beta=0\arcminf25$ around the center of the data
field. Fig.~\ref{fig:2} shows that the quantities (\ref{eq:3.2}) vary
considerably between realizations, owing to the broad distribution of
intrinsic source properties. When averaged over 50 realizations
(Figs.~\ref{fig:2}d,e), their mean values are very close to unity, and
their 1-$\sigma$ variations are a good indicator for the expected
values in true reconstructions.

Even when the exact deflection potential is used, the resulting mass
distribution $\kappa$ will deviate from the true surface mass density
because $\kappa$ is calculated from $\psi$ with finite
differencing. For the mass distribution shown in Fig.~\ref{fig:1},
$\kappa$ from finite differencing deviates from the true $\kappa$ by
$\lesssim0.03$ everywhere, with the largest deviations occurring at
the two mass peaks. The grid is too coarse for a more accurate
calculation of the second-order derivatives. Since the deviations are
sufficiently small (i.e.~much smaller than the expected accuracy that
we can hope to achieve from our reconstruction), we have chosen not to
further refine the grid. It should be noted that the method by Bridle
et al.\ (1998) suffers from the same, or worse, inaccuracies, because
there the shear is calculated from the surface mass density by
integrating over a coarse grid.

Finally, Bridle et al.\ (1998) calculated the covariance matrix for
the resulting mass distribution, which we do not repeat here. One must
take into account, though, that the error estimates of the resulting
mass reconstruction are strongly correlated, because the shear depends
non-locally on $\kappa$.

\section{Results}

\subsection{ML reconstructions without magnification information}

We first neglect magnification information, i.e.~we minimize
(\ref{eq:2.13}). For the mass model in Fig.~\ref{fig:1}, mass
reconstructions for 50 realizations of the galaxy population were
performed. For each realization, 1000 iterations were taken, in each
case using the noise-filter reconstruction of Seitz \& Schneider
(1996) as initial potential. The number of iterations was chosen to
produce a stable result for all realizations. The actually required
number can be substantially smaller in individual cases. Typically,
1000 iterations take approximately 30 minutes on an IBM 590
workstation. After every 20 iterations, the prior was changed to the
current mass distribution, smoothed by a Gaussian of width
$\theta_{\rm sm}=0\arcminf2$ for the first 600 iteration steps, and
$\theta_{\rm sm}=0\arcminf15$ afterwards. A somewhat larger smoothing
length at the beginning leads to faster convergence, but produces
artifacts due the finite region over which the integral in
(\ref{eq:2.3}) is performed. Note that $\theta_{\rm sm}$ is of the
same size as the grid cells. In order to avoid excessive fine-tuning
for these simulations, the regularization parameter $\eta$ was fixed
to $\eta=30$ for all 50 realizations. Of course $\eta$ could be
changed to finally achieve $G_\chi=1$ for each individual
realization. The current choice of $\eta$ was made to achieve
$G_\chi\approx1$ for all realizations (see Fig.~\ref{fig:3}
below). The final 20 iteration steps were performed with a value of
$\eta$ large enough to make the mass distribution follow the prior
very closely, and which yields a smoothing of the mass reconstruction
on a scale of $\theta_{\rm sm}$.

\begin{figure*}
  \resizebox{\bsize}{!}{\includegraphics{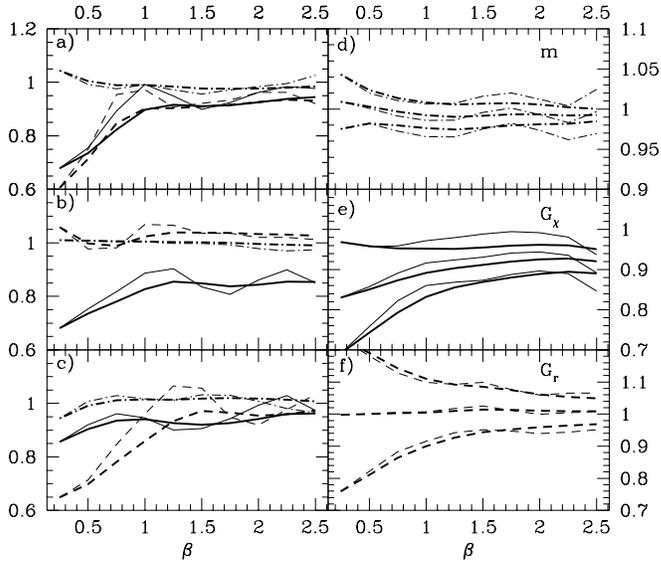}}
  \hfill
  \parbox[b]{\csize}{
\caption{Panels (a) through (c) show the chi-square quantities
  (\ref{eq:3.2}), together with the mass ratios (\ref{eq:4.1}) for the
  reconstructions obtained from three different realizations of the
  galaxy population. Heavy solid curves: $G_\chi(<\beta)$; light solid
  curves: $G_\chi(\beta)$; heavy dashed curves: $G_r(<\beta)$; light
  dashed curves: $G_r(\beta)$; heavy dash-dotted curves: $m(<\beta)$;
  light dash-dotted curves: $m(\beta)$. In panels (d) through (f), the
  average of these quantities from 50 realizations are displayed
  together with their 1-$\sigma$ range (shown by the same line
  types). For these simulations, the regularization parameter was kept
  constant, i.e.~not adjusted for each simulation to yield
  $G_\chi\approx 1$. No magnification information was used}
\label{fig:3}}
\end{figure*}

Since no magnification information was used in these simulations, the
resulting deflection potential (and thus the mass distribution) is
determined only up to the mass-sheet transformation (\ref{eq:2.6}). In
order to compare the reconstructions with the input model, each mass
map was transformed such that the total mass inside $\U$ agreed with
the true total mass. In Fig.~\ref{fig:3}, we show the quantities
(\ref{eq:3.2}) for three different realizations of the galaxy
population, together with the ratios
\begin{eqnarray}
  m(\beta)  &:=& {M(\beta)\over M_{\rm true}(\beta)}\; ,\nonumber\\
  m(<\beta) &:=& {M(<\beta)\over M_{\rm true}(<\beta)}
\label{eq:4.1}
\end{eqnarray}
of the reconstructed mass inside rings and circles, relative to the
true mass distribution. As can be seen, these mass ratios are always
very close to unity, which means that the mass maps are reconstructed
with high accuracy (up to the mass-sheet degeneracy). The dispersion
of $m(\beta)$ about unity is less than 5\%, and the mean of $m(\beta)$
over 50 realizations is astonishingly flat. There is no indication
that the mass at the center or in the outer parts is systematically
over- or underestimated. The choice of the regularization parameter
results in $G_\chi$ being slightly smaller than unity on average,
though with substantial variation from case to case. Evidently,
$G_\chi(\beta)$ is significantly smaller in the inner part than in the
outer part, an effect also seen in Fig.~\ref{fig:2} where the true
mass distribution was considered.  This is due to the fact that the
ellipticity distribution after lensing is not really a Gaussian, and
the deviation from this assumed functional form becomes larger for
larger values of the reduced shear, i.e.\ closer to the center of
$\U$. Whereas there is no fundamental difficulty in replacing the
Gaussian with a more accurate probability distribution, the simple
form for $G_\chi$ is computationally convenient and seems to be
sufficiently accurate for the mass reconstructions, as seen from the
dash-dotted curves in Fig.~\ref{fig:3}. The deviation of $G_r(\beta)$
from unity can be substantial in individual reconstructions, but the
mean over all realizations is very close to unity.

At the edge of the data field, two systematic effects become visible
in the mean quantities plotted in Figs.~\ref{fig:3}d--f (and also by
looking at the 2-d distribution over $\U$): The value of
$G_\chi(\beta)$ shows a small but significant decrease, and $\kappa$
is slightly too large near the boundary. The first of these effects
can be understood by considering the number of galaxies for which the
shear estimate is affected if the value of $\psi$ is changed at a grid
point. If that grid point is located in the inner part of $\U$, the
estimates of $\gamma$ for galaxies within the neighboring 16 grid
cells are affected. This number decreases for points near the boundary
of $\U$, so that $\psi$ there is less constrained by the measured
image ellipticities. This implies that at the boundary, it is easier
to ``fit the noise'' caused by the intrinsic ellipticity
distribution. The slightly too large $\kappa$ near the boundary is due
to the prior. The prior is obtained from local averaging of the
current mass distribution. If the mass distribution decreases
outwards, the local mean value which can only be taken from the grid
points within $\U$ will be slightly too large at the boundary, which
explains why the method presented here is slightly biasing the mass
map at the boundary. In actual applications, a strip of width
$\theta_{\rm sm}$ can be ignored in the analysis of the mass
distribution if this bias is a worry. However, its amplitude is very
small, and it can probably be safely ignored in most situations
compared to the stochastic errors. Alternatively, one can use a mild
extrapolation of the smoothed mass distribution to obtain an estimate
of the smoothed $\kappa$ values on the boundary less affected by this
bias. In the case of our mass model, a simple fix (instead of a more
elaborated extrapolation) can be obtained by decreasing $\kappa$ by
$\Delta\kappa=0.0015$ on all boundary points, which practically
eliminates the bias. We use this simple fix to the simulations
discussed in the next subsection. We further point out that the
amplitude of this mass bias can also be checked with real data, by
generating artificial data sets from the reconstructed mass
distribution and by performing reconstructions for those in the same
way in which the original mass reconstruction was obtained.

\begin{figure}[ht]
  \centerline{%
    \resizebox{\asize}{!}{\includegraphics{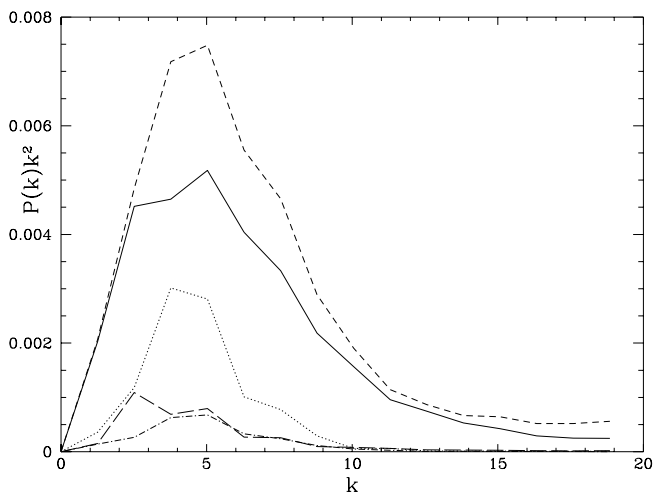}}}
\caption{Power spectra $k^2 P(k)$ of the difference between the values
  of $K(\vc\theta)=\ln[1-\kappa(\vc\theta)]$ from reconstructed mass
  distributions and the true mass distribution (see Seitz \& Schneider
  1996 for a definition of $P(k)$ and its practical calculation). The
  solid curve corresponds to $K_{\rm ML}$, the short-dashed curve to
  $K_{\rm NF}$. In addition, we plot the power spectra of the
  difference between the ensemble-averaged mass maps, both from the ML
  method (long-dashed curve) and the noise-filter reconstruction
  (dotted curve). For comparison, the dash-dotted curve shows the
  power spectrum of the difference obtained from the
  finite-differencing error, mentioned at the end of Sect.\ 3}
\label{fig:4}
\end{figure}

The mass-sheet transformation (\ref{eq:2.6}) allows to determine the
quantity $K(\vc\theta)=\ln[1-\kappa(\vc\theta)]$ up to an additive
constant. As in Seitz \& Schneider (1996, 1998; SK), we perform a
power-spectrum analysis of the difference between $K_{\rm ML}$ and
$K_{\rm true}$, i.e.~between the $K$ from the ML reconstructions and
from the true mass distribution. Note that this power spectrum for
$k\ne0$ is independent of the transformation (\ref{eq:2.6}). In
Fig.~\ref{fig:4}, we compare the power spectrum obtained from our 50
realizations to the power spectrum obtained from the difference
between $K_{\rm NF}$ and $K_{\rm true}$, the former one corresponding
to the noise-filter mass reconstruction of Seitz \& Schneider (1996),
using the same realizations of the galaxy population. As can easily be
seen, the ML reconstruction yields substantially lower values for
$P(k)$ than the noise-filter reconstruction.

For illuminating the main reason for this difference, we also
calculate the power spectra of the difference between $\ave{K_{\rm
ML}}$, $\ave{K_{\rm NF}}$ and $K_{\rm true}$, the former two being the
mean reconstructed mass distribution averaged over all 50
realizations. The difference between these two power spectra is very
large, with the power of $\ave{K_{\rm NF}}-K_{\rm true}$ being much
larger than for the ML reconstructions. Almost the entire difference
comes from the inability of the noise-filter method to resolve the
core of the two mass peaks in the model appropriately because of the
unavoidable smoothing. Note that the smoothing scale $\Delta\theta$ in
the noise-filter reconstruction is much larger than the smoothing
scale $\theta_{\rm sm}$ for the prior in the ML method, the latter
being of the same order as the size of a grid cell, or the mean
separation between two background galaxy images, which in any case is
smallest scale one can hope to resolve. In fact, it seems that the
power spectrum of $K_{\rm NF}-K_{\rm true}$ is approximately the sum
of the power spectra of $K_{\rm ML}-K_{\rm true}$ and $\ave{K_{\rm
NF}}-K_{\rm true}$. In other words, the error of the noise-filter
reconstruction is due to a systematic error (from smoothing, although
it is adaptive smoothing as in Seitz \& Schneider 1996), and a
stochastic error which is approximately the same as that for the ML
reconstruction.

The power spectrum of the error resulting from finite differencing is
shown as the dash-dotted curve in Fig.~\ref{fig:4} and is seen to be
about the same as that of $\ave{K_{\rm ML}}-K_{\rm true}$. I.e., the
largest part of the error of the mean mass map in the ML
reconstruction comes from finite differencing; only for $k\le3$ is the
latter error larger than that from finite differencing. This little
power excess is due to the aforementioned slight bias at the boundary
of $\U$.

To conclude this subsection, we have shown that the ML method yields
more accurate mass reconstructions than the noise-filter method,
whereby the largest difference is due to the better angular resolution
of the ML method in those parts of the data field where the data
require more local structure, whereas the noise-filter method -- like
all direct methods developed so far -- uses a fixed smoothing scale,
or at best an ad-hoc choice for the change of smoothing scale with the
signal, such as in Seitz et al.\ (1996, and references therein). The
ML mass distributions are slightly biased at the boundary of the data
field, and the amplitude of this bias depends on the mass
distribution.

\subsection{ML reconstructions with magnification information}

For the 50 realizations of the background galaxies, we also performed
ML mass reconstructions including magnification information,
i.e.~minimizing (\ref{eq:2.16}). Each reconstruction proceeded as
described in the beginning of Sect.\ 4.1, except that the
regularization parameter was chosen to be $\eta=60$ instead of
30. This is necessary to give the regularization the same weight as
before, because the chi-square part of $E$ is now effectively
doubled. The number of iteration steps was slightly increased to
1200. Anticipating that the additional magnification information would
allow a slightly higher angular resolution of the mass map, we
somewhat reduced the smoothing scale for the prior to $\theta_{\rm
sm}=0\arcminf1$.

\begin{figure*}
  \resizebox{\bsize}{!}{\includegraphics{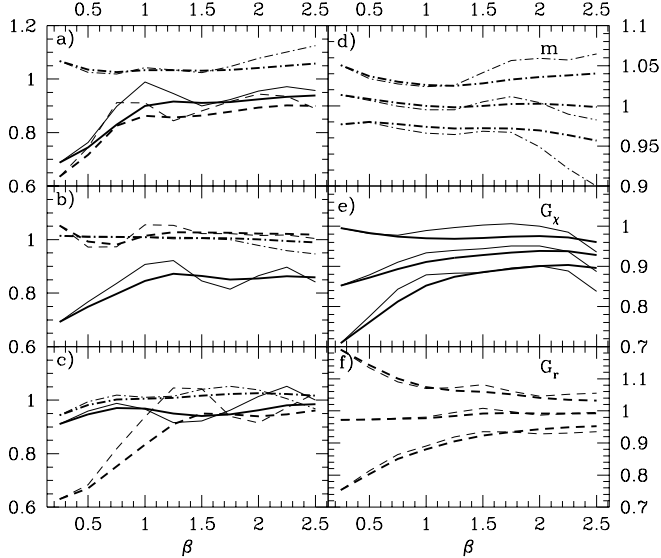}}
  \hfill
  \parbox[b]{\csize}{
\caption{In analogy to Fig.~\ref{fig:3}, we plot in panels (a) through
  (c) the chi-square contributions (\ref{eq:3.2}), together with the
  mass ratios (\ref{eq:4.1}), for the ML-reconstructions obtained from
  three different realizations of the galaxy population, now using
  magnification information. Heavy solid curves: $G_\chi(<\beta)$,
  light solid curves: $G_\chi(\beta)$; heavy dashed curves:
  $G_r(<\beta)$; light dashed curves: $G_r(\beta)$; heavy dash-dotted
  curves: $m(<\beta)$; light dash-dotted curves: $m(\beta)$. In panels
  (d) through (f), the average of these quantities over 50
  realizations is displayed together with their 1-$\sigma$ range. For
  these simulations, the regularization parameter was kept constant,
  i.e.~not adjusted for each simulation to achieve $G_\chi\approx 1$}
\label{fig:5}}
\end{figure*}

In analogy to Fig.~\ref{fig:3}, we plot in Fig.~\ref{fig:5} the
quantities (\ref{eq:3.2}) and (\ref{eq:4.2}) for three different
realizations of the galaxy population, their average over 50
realizations, and their dispersion. Comparing the two figures, it
seems that including magnification information does not change the
curves significantly, apart from the fact that the mass-sheet
transformation is now obsolete. Magnification information therefore
mainly affects the normalization of the mass map, but does not provide
significant information on its shape.

\begin{figure}[ht]
  \centerline{%
    \resizebox{\asize}{!}{\includegraphics{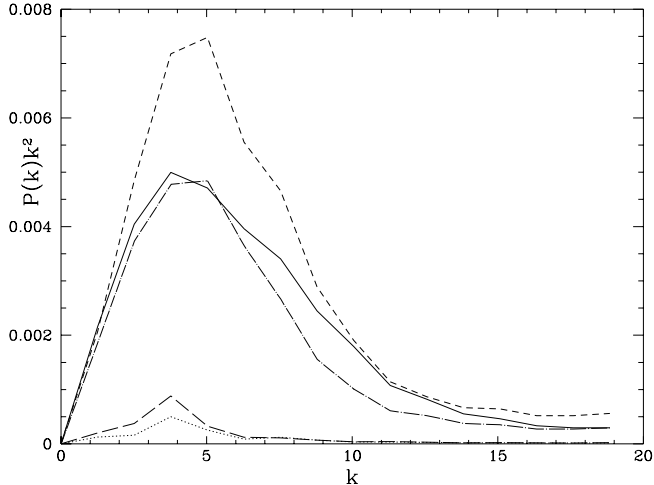}}}
\caption{Power spectra $k^2 P(k)$ of the difference between the
  reconstructed mass distributions
  $K(\vc\theta)=\ln[1-\kappa(\vc\theta)]$ and the true mass
  distribution. The solid and long-dash-dotted curves show $K_{\rm
  ML}$ obtained without and with magnification information,
  respectively.  For comparison, the short-dashed curve shows $K_{\rm
  NF}$. In addition, we plot the power spectra of the difference
  between the ensemble-averaged mass distributions, both from the ML
  method without (long-dashed curve) and with (dotted curve)
  magnification information}
\label{fig:6}
\end{figure}

To further illustrate this point, we plot in Fig.~\ref{fig:6} the
power spectra of the difference between reconstructed mass maps
$K_{\rm ML}$ and the model mass distribution $K_{\rm true}$, similar
to Fig.~\ref{fig:4}. The difference in the power spectra between
reconstructions with and without magnification information is much
smaller than their difference to the noise-filter mass
reconstructions, at least for small values of $k\lesssim7$. For larger
$k$, the difference becomes somewhat larger, but this may simply be
due to the different choice of the regularization parameter which
affects the small-scale noise. In our model, these small scales are
only noise because the mass model is relatively smooth. The power
spectra of the ensemble-averaged mass maps show that the spectrum for
simulations with magnification information is somewhat lower. This is
probably due to the fact that the magnification information leads to a
slightly better angular resolution in the core of the two mass
concentrations.

Our simulation results therefore allow to conclude that magnification
information does not contribute much information on the {\em shape\/}
of the mass distribution compared to shear information. For estimating
the relative accuracy of the mass determination, one can then assume
that the mass map is known from the galaxy shapes up to the mass-sheet
transformation. Let $\kappa$ be the true mass distribution and
$\kappa(\lambda)$ the distribution transformed with
(\ref{eq:2.6}). The magnification information can then be used to
determine $\lambda$ such as to best fit the galaxy sizes, i.e.~to
minimize $G_r$. Noting that (\ref{eq:2.6}) transforms magnifications
like $\mu\to\mu/\lambda^2$, the best-fitting $\lambda$ is
\begin{equation}
  \ln\lambda = -{1\over N_{\rm g}}\sum_{k=1}^{N_{\rm g}}\,
  \eckk{\ln r(k)-\abs{\mu(\vc\theta_k)}^{1/2}}\;.
\label{eq:4.2}
\end{equation}
Since the distribution of the $\ln r$ is Gaussian, the values of
$\ln\lambda$ for different realizations of the galaxy population
follow a Gaussian probability distribution with zero mean and
dispersion $\sigma_\lambda=\sigma_r/\sqrt{N_{\rm g}}$. Whereas the
distribution of $\ln\lambda$ is symmetric about zero, the mean of
$\lambda$ is not unity. The ratio
\begin{equation}
  m(\lambda) = {M(\lambda)\over M} = 
  {\ave{\kappa(\lambda)}\over\ave{\kappa}} = 
  {1\over\ave{\kappa}} + \lambda\rund{1-{1\over\ave{\kappa}}}
\label{eq:4.3}
\end{equation}
of the estimated mass to the true mass has a median of unity, but a
slightly smaller mean,
\begin{eqnarray}
  \ave{m} &=& \int_{-\infty}^\infty\d\ln\lambda\,
  p(\ln\lambda)\,m(\ln\lambda) \nonumber\\
  &\approx& 1-\rund{{1\over \ave{\kappa}}-1}\,
  {\sigma_r^2 \over 2 N_{\rm g}}\;.
\label{eq:4.4}
\end{eqnarray}
The deviation of $\ave{m}$ from unity is very small: For $N_{\rm
g}=1250$ and $\sigma_r=0.5$, and a mass model with
$\ave{\kappa}\approx 0.2$, we find $\ave{m}\approx0.9996$. Similarly,
the dispersion of $m$ is
\begin{equation}
  \sigma_m=\sqrt{\ave{m^2}-\ave{m}^2}\approx\abs{{1\over
  \ave{\kappa}}-1} {\sigma_r\over \sqrt{N_{\rm g}}}\;,
\label{eq:4.5}
\end{equation}
where we have used $\sigma_r/ \sqrt{N_{\rm g}}\ll 1$ in the
approximations. For the same parameters as before, $\sigma_m=0.057$.
We therefore expect that our mass model can be reconstructed with an
accuracy of about 6\% if the shape of the mass distribution is
sufficiently constrained by image shapes. Fig.~\ref{fig:5}d shows that
this estimate in very good agreement with the mass estimate obtained
from the ML reconstruction using the magnification information.

\begin{figure}[ht]
  \centerline{%
    \resizebox{\asize}{!}{\includegraphics{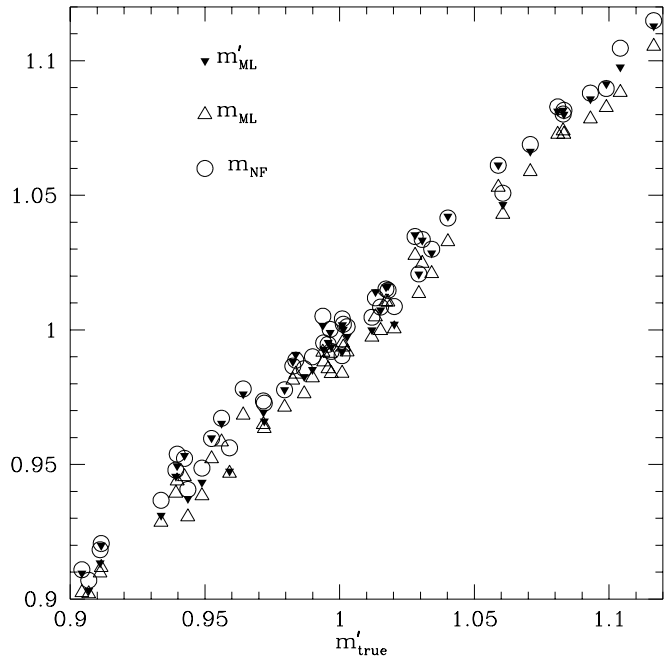}}}
\caption{For 50 realizations of the galaxy population, we estimate the
  total mass inside $\U$ and compare it to the true mass. Defining
  $m:=M/M_{\rm true}$, we plot (i) the mass ratio $m_{\rm ML}$
  obtained from ML reconstructions including magnification
  information, (ii) the ratio $m'_{\rm ML}$ obtained from ML
  reconstructions without magnification information, but mass-sheet
  transformed according to (\ref{eq:4.2}), and (iii) the mass ratio
  $m_{\rm NF}$ obtained from noise-filter reconstructions, transformed
  according to (\ref{eq:4.2}), against (iv) the ratio $m'_{\rm true}$
  obtained from the true shape of the mass distribution, also
  invariance transformed according to (\ref{eq:4.2}). The means and
  dispersions of the four mass ratios are: (i) $\ave{m_{\rm
  ML}}=0.994$, $\sigma_m=0.051$; (ii) $\ave{m'_{\rm ML}}=1.000$,
  $\sigma_m=0.052$; (iii) $\ave{m_{\rm NF}}=1.002$, $\sigma_m=0.051$;
  and (iv) $\ave{m'_{\rm true}}=1.001$, $\sigma_m=0.054$. Within the
  accuracy allowed by 50 realizations, they all agree with $\ave{m}=1$
  and $\sigma_m\approx 0.052$, very much in agreement with the
  estimate (\ref{eq:4.5})}
\label{fig:7}
\end{figure}

To elaborate on this point, we plot in Fig.~\ref{fig:7} the ratios of
four different mass estimates to the true mass. These are: the mass
from ML reconstructions (i) with and (ii) without magnification
information, but transformed according to (\ref{eq:4.2}) to best fit
the source sizes; (iii) mass-sheet-transformed noise-filter
reconstructions; and (iv) the mass-sheet-transformed true density
distribution. Obviously, the scatter of the mass estimates across the
four different methods is very much smaller than that caused by
different realizations of the source population. As much as 50
realizations allow to conclude, all four methods yield an unbiased
estimate of the total mass and have approximately the same dispersion,
close to the theoretical expectation discussed above.

We investigated how the quality of mass estimates is affected by the
spatial resolution. The noise-filter reconstructions have a lower
resolution than the ML reconstructions, although our adaptive
smoothing makes them satisfactorily resolve the mass peaks. We
performed noise-filter reconstructions with fixed smoothing scales of
$0\arcminf3$ and $0\arcminf5$. In those cases, the peaks are fairly
much smoothed out. Applying then (\ref{eq:4.2}) to match image sizes,
we found that the total mass within $\U$ is systematically
overestimated, by $\sim 1$\% and $\sim 3$\%, respectively. This is
because the smoothing of the map renders the model magnification too
low to match the observed high magnifications close to mass peaks, for
which the invariance transformation tends to compensate with a
slightly higher mass. This bias also affects the ML reconstruction of
BNSS because of missing resolution in the cluster center. However, it
is sufficiently small to be not of much practical importance for
reconstructions which can be expected from real data in the near
future, but it should be kept in mind if a large sample of clusters is
investigated statistically.

\section{Discussion and conclusions}

We presented a new method for reconstructing projected mass
distributions of galaxy clusters. The method uses image distortions of
background galaxies and their size as a function of surface
brightness. Our entropy-regularized ML method (Seitz 1997) is a
further development of previously published inverse methods for the
mass reconstruction. In particular, we describe the lens by its
deflection potential $\psi$ as suggested by BNSS. This is of major
importance, for two reasons. First, if the surface mass density
$\kappa$ on a finite field $\U$ is used to describe the lens, the
shear on $\U$ is incompletely specified by the model because the mass
distribution outside $\U$ can contribute to the shear. Second, the
shear at the position of any galaxy depends only locally on $\psi$,
which allows a much faster minimization algorithm for a given number
of grid points.

We regularize the method by an entropy term as suggested by Bridle et
al.\ (1998), but additionally adapt the prior to the current model of
the mass distribution. This `moving prior' (Lucy 1994) allows a
considerably higher resolution of mass peaks. The spatial resolution
of the entropy-regularized ML method adapts itself to the strength of
the lensing signal, producing mass distributions which are as smooth
as possible, and as structured as the data require. In that respect,
our method differs from that of BNSS and SK. We showed that the ML
method is superior to the noise-filter method (Seitz \& Schneider
1996) which was the most accurate of the presently known direct
inversion methods (Seitz \& Schneider 1996, 1998; SK; Lombardi \&
Bertin 1998).

Obviously, the method described here is not restricted to rectangular
data fields, but can easily be adapted to any geometry of $\U$ by
covering $\U$ with quadratic grid cells, and adding a boundary of grid
points for $\psi$ -- the rest is only a matter of
labeling. Furthermore, we note that observational errors can be
incorporated into the likelihood function. For example, if the
measurement error of the ellipticity $\chi$ is $\sigma_{\rm obs}$, one
can replace $\sigma_\chi^2$ in (\ref{eq:2.13} and \ref{eq:2.14}) by
$\sigma_\chi^2+\sigma_{\rm obs}^2$.

In contrast to the direct inversion methods, all of which are variants
and generalizations of the original Kaiser \& Squires (1993) method,
the inverse methods allow to include additional information on top of
the shear measured through image ellipticities. We demonstrated this
here by adding magnification information derived from image sizes at
given surface brightness, as discussed by Bartelmann \& Narayan
(1995). However, we could equally well use the change of number counts
due to magnification bias (Broadhurst et al.\ 1995) as an additional
constraint. In that case, if the number counts of a certain
(e.g.~color-selected) galaxy population have a cumulative slope of
$-\beta$, the expected number density of background galaxies at a
position $\vc\theta$ is
$n(\vc\theta)=n_0\abs{\mu(\vc\theta)}^{\beta-1}$, where $n_0$ are the
counts at the same flux limit in the absence of lensing. Assuming that
galaxies are intrinsically randomly distributed, the probability of
having $N$ galaxies within $\U$ is a Poisson distribution
$P_N(\ave{N})$ with
\begin{equation}
  \ave{N} = n_0\int_\U\,\d^2\theta\,
  \abs{\mu(\vc\theta)}^{\beta-1}\;.
\label{eq:5.1}
\end{equation}
Consequently, the likelihood function could be augmented by a factor
\begin{equation}
  \L_\mu = P_N(\ave{N})\prod_{k=1}^N\,
  \abs{\mu(\vc\theta_k)}^{\beta-1}\;.
\label{eq:5.2}
\end{equation}
If galaxy clustering is important, the likelihood $\L_\mu$ cannot be
written as a simple product over individual galaxies, but the joint
probability distributions must be taken into account. The contribution
of clustering effects to the likelihood function is somewhat
uncertain, because an approximate expression has to be used due to
lack of knowledge on the $N$-point correlation functions (see
Broadhurst et al.\ 1995 for further discussion).

Perhaps the most promising generalization of our method is the
inclusion of strong lensing constraints. Since giant arcs and multiple
images of background galaxies provide (nearly) exact constraints on
the lens mass distribution, it is highly desirable to include them
into a mass reconstruction. The obvious way to do this would be to
augment the function $E$ by a term which measures the degree to which
multiple images of the same source are mapped back to the same
position in the source plane. In addition, the surface brightness
profile of multiple images of extended sources can be incorporated,
e.g.~in a similar manner as the spatially resolved multiple arc in the
cluster Cl0024+16 (Colley at al.\ 1996).

In some of the observed clusters, the lensing effects of individual
galaxies are visible, in particular through deformations of giant
arcs. Some of the most prominent examples are the triple arc in
0024+16 (Kassiola et al.\ 1992), the multiple arc systems in A~2218
(Kneib et al.\ 1996), and the distortion of the curvature in the arc
of the galaxy cB58 in MS1512+36 (Seitz et al.\ 1998). But even weaker
lensing effects of individual (cluster) galaxies can be detected using
a combination of cluster mass reconstruction and galaxy-galaxy lensing
techniques. By adding two free parameters to the lens model, such as
the mass-to-light ratio of cluster galaxies and their characteristic
spatial extent, the size of halos of cluster galaxies can be
investigated (Natarayan et al.\ 1997; Geiger \& Schneider 1998).

\section*{Acknowledgements} 

This work was supported by the ``Sonderforschungsbereich 375-95''
f\"ur Astro-Teilchenphysik der Deutschen Forschungsgemeinschaft.


\appendix

\section{Notes on the implementation}

We outline here how the function $E$ -- see (\ref{eq:2.13}) and
(\ref{eq:2.16}) -- can be calculated from the deflection potential
$\wave{\psi}$ at the grid points. We first note that $E$ can be easily
calculated in terms of the shear $\gamma$ and the surface mass density
$\kappa$ at the position of the galaxies $\vc\theta_k$, and in terms
of $\kappa$ at the grid points, which enters into the regularization
term $\R$.

Let $\psi_\alpha$ label the deflection potential at the $N_{\rm dim}$
grid points where $\psi$ is varied. If no magnification information is
used, $N_{\rm dim}=(N_x+2)(N_y+2)-8$, and $N_{\rm
dim}=(N_x+2)(N_y+2)-7$ otherwise. The components of $\gamma$ and
$\kappa$ at the $N_xN_y$ grid points are obtained by second-order
finite differencing, according to (\ref{eq:2.1}) and (\ref{eq:2.2}).
For calculating $\gamma_2$ at the grid corners, we used the
finite-difference formula (25.3.27) from Abramowitz \& Stegun (1984)
so that $\psi$ at the corner points of the extended grid does not
enter $E$. This can be important because the corner points do not
enter the entropy term, so that they are not regularized. If no galaxy
happens to lie in the corner grid cells, these points would be
unconstrained. Finite differencing being a linear operation, we can
write
\begin{equation}
  \kappa_\mu = D_{\mu\alpha}\psi_\alpha + 
  d_\mu\;,\;1\le \mu\le N_xN_y\;.
\label{eq:A1}
\end{equation}
Summation over $\alpha$, $1\le\alpha\le N_{\rm dim}$, is implied, and
the array $d_\mu$ accounts for the contribution to $\kappa_\mu$ from
the grid points where $\psi$ is held constant. For each galaxy
position $\vc\theta_k$, the values of $\kappa$ and $\gamma$ are
obtained from bilinear interpolation from the closest grid points.
Since interpolation is again a linear operation, we can write
\begin{eqnarray}
  \gamma_1(\vc\theta_k) &=& A_{k\alpha}\psi_\alpha+a_k\;, \nonumber\\
  \gamma_2(\vc\theta_k) &=& B_{k\alpha}\psi_\alpha+b_k\;, \nonumber\\
  \kappa(\vc\theta_k)   &=& C_{k\alpha}\psi_\alpha+c_k\;,
\label{eq:A2}
\end{eqnarray}
for $1\le k\le N_{\rm g}$. The arrays $a$, $b$, and $c$ again arise
from the contributions of the grid points where $\psi$ is held fixed.
If magnification information is used, $\psi$ is kept fixed at three
grid points, where $\psi=0$ can be chosen without loss of generality;
then, $a_k=b_k=c_k=0=\d_\mu$. If magnification information is
neglected, at least one of the four constant values of $\psi$ has to
differ from zero, and $a$, $b$, $c$, and $d$ are not identically zero,
although most of their elements vanish. The same holds for the
elements of the matrices $A$, $B$, $C$, and $D$. Whereas the dimension
of these matrices is large, only about $12N_{\rm g}$ elements are
different from zero. Thus, we store these matrices using the
row-indexed sparse storage mode, as described in Press et al.\ (1992,
Sect.\ 2.7). Note that these matrices and `vectors' have to be
calculated only once, since they depend only on the number of grid
points and the location of the galaxies. The sparseness of the
matrices is one of the advantages to work in terms of the deflection
potential rather than the surface mass density.

For the calculation of $\partial E/\partial\psi_\alpha$, we note that
for any function $f_k$ at $\vc\theta_k$,
\begin{equation}
  {\partial f_k\over \partial \psi_\alpha} =
  {\partial f_k\over \gamma_1(\vc\theta_k)}A_{k\alpha} +
  {\partial f_k\over \gamma_2(\vc\theta_k)}B_{k\alpha} +
  {\partial f_k\over \kappa(\vc\theta_k)}C_{k\alpha}\;.
\label{eq:A3}
\end{equation}
For the regularization term, we find
\begin{eqnarray}
  {\partial \R\over \partial \psi_{\alpha}} &=&
  -\eck{\sum_{\mu=1}^{N_x N_y} \kappa_\mu}^{-1} \nonumber\\
  &\times& \sum_{\mu=1}^{N_x N_y}\,
  \ln\rund{\hat\kappa_\mu\over b_\mu}\,
  \eck{D_{\mu\alpha}-\hat\kappa_\mu\sum_{\nu=1}^{N_x N_y}\,
  D_{\nu\alpha}}\;.
\label{eq:A4}
\end{eqnarray}
Using these relations, the function $E$ and its derivative with
respect to the $\psi_\alpha$ are easily calculated.

An additional complication arises for a critical lens. As can be seen
from the explicit form of the terms of $E$, the likelihood function
attains singularities if a galaxy image happens to lie on a critical
curve where $\abs{\delta}=1$ and $\mu^{-1}=0$. This means that the
minimization procedure cannot modify the model such that critical
curves move over galaxy positions. Since the initial guess of $\psi$
will certainly not be accurate enough that such crossings can be
avoided, we have to modify the function $E$ appropriately. The
necessary modification of $E$ can be achieved by replacing $\abs{\mu}$
by $\eck{\mu^{-2}+\eps^2}^{-1/2}$, and $\sigma_\chi$ by
$\sigma_\chi+\eta$, where $\eps$ and $\eta$ are two small quantities.
These replacements leave $E$ finite if a galaxy image is placed on a
critical curve. The minimization then proceeds by setting $\eps$ and
$\eta$ to about 0.1 at the beginning of the minimization, and then
slowly decreasing them in later iteration steps. This leads to
convergence without additional problems. In addition, as was true for
the direct inversions, by considering a broad redshift distribution of
the background galaxies (Seitz \& Schneider 1997), the singularities
connected with critical curves are avoided (Geiger \& Schneider 1998).


\begin{thebibliography}{}

\bibitem{ref:1} Abramowitz, M., Stegun, I.\ 1984, ``Handbook of
  Mathematical Functions'', Harri Deutsch Verlag
\bibitem{ref:2} Bartelmann, M. \ 1995, A\&A, 303, 643
\bibitem{ref:3} Bartelmann, M., Narayan, R.\ 1995, ApJ, 451, 60
\bibitem{ref:4} Bartelmann, M., Narayan, R., Seitz, S., Schneider, P.\
  1996, ApJ, 464, L115 (BNSS)
\bibitem{ref:5} Blandford, R.D., Saust, A.B., Brainerd, T.G.,
  Villumsen, J.V.\ 1991, MNRAS 251, 600
\bibitem{ref:6} Bridle, S.L., Hobson, M.P., Lasenby, A.N., Saunders,
  R., 1998, astro-ph/9802159
\bibitem{ref:7} Broadhurst, T.J., Taylor, A.N. Peacock, J.A.\ 1995,
  ApJ, 438, 49
\bibitem{ref:8} Colley, W.N., Tyson, J.A., Turner, E.L.\ 1996, ApJ
  461, L83
\bibitem{ref:9} Fort, B., Mellier, Y., Dantel-Fort, M.\ 1997, A\&A,
  321, 353
\bibitem{ref:10} Geiger, B., Schneider, P.\ 1998, in preparation
\bibitem{ref:11} Gorenstein, M.V., Falco, E.E., Shapiro, I.I.\ 1988,
  ApJ, 327, 693
\bibitem{ref:12} Kaiser, N., Squires, G.\ 1993, ApJ, 404, 441
\bibitem{ref:13} Kaiser, N., Squires, G., Fahlmann, G.G., Woods, D.,
  Broadhurst, T.\ 1994, astro-ph/9411029
\bibitem{ref:14} Kaiser, N.\ 1995, ApJ, 493, L1
\bibitem{ref:15} Kaiser, N., Squires, G., Broadhurst, T.\ 1995,
  ApJ,449, 460
\bibitem{ref:16} Kassiola, A., Kovner, I., Fort, B.  1992, APJ, 400,
  41
\bibitem{ref:17} Kneib, J.-P., Ellis, R.S., Smail, I., Couch, W.J.,
  Sharples, R.M.\ 1996, ApJ 471, 643
\bibitem{ref:18} Kolatt, T.S., Bartelmann, M.\ 1998, MNRAS, in press
\bibitem{ref:19} Lombardi, M., Bertin, G.\ 1998, astro-ph/9801244
\bibitem{ref:20} Lucy, L.\ 1994, A\&A 289, 983
\bibitem{ref:21} Narayan, R., Nityananda, R.\ 1986, ARA\&A 24, 127
\bibitem{ref:22} Natarayan, P., Kneib, J.-P., Smail, I., Ellis, R.S.\
  1997, astro-ph/9706129
\bibitem{ref:23} Press, W.H., Teukolsky, S.A., Vetterling, W.T.,
  Flannery, B.P.\ 1992, Numerical Recipes. Cambridge (Cambridge
  University Press)
\bibitem{ref:24} Schneider, P.\ 1995, A\&A, 302, 639
\bibitem{ref:25} Schneider, P., Bartelmann, M.\ 1997, MNRAS 286, 673
\bibitem{ref:26} Schneider, P., Seitz, C.\ 1995, A\&A, 294, 411
\bibitem{ref:27} Seitz, C., Schneider, P.\ 1995, A\&A, 297, 287
\bibitem{ref:28} Seitz, C., Kneib, J.P., Schneider, P., Seitz, S.\
  1996 A\&A, 314, 707
\bibitem{ref:29} Seitz, C., Schneider, P.\ 1997, A\&A, 318, 617
\bibitem{ref:30} Seitz, S., Schneider, P.\ 1996, A\&A, 305, 383
\bibitem{ref:31} Seitz, S.\ 1997, ``Untersuchungen zum schwachen
  Linseneffekt auf Quasare und Galaxien''. Ph.D.\ Dissertation (in
  German), Ludwig-Maximilians-Universit\"at M\"unchen.
\bibitem{ref:32} Seitz, S., Saglia, R.P., Bender, R., Hopp, U.,
  Belloni, P., Ziegler, B.\ 1998, MNRAS, in press
\bibitem{ref:33} Seitz, S., Schneider, P.\ 1998, A\&A submitted,
  astro-ph/9802051
\bibitem{ref:34} Squires, G., Kaiser, N.\ 1996, ApJ, 473, 65 (SK)

\end{thebibliography}
\end{document}